\documentclass[prl,twocolumn,showpacs]{revtex4}
\usepackage{graphicx}
\catcode`\@=11
\def\simle{\mathrel{\mathpalette\@versim<}}   % < over \sim
\def\simge{\mathrel{\mathpalette\@versim>}}   % > over \sim
\def\@versim#1#2{\lower2.5pt\vbox{\baselineskip0pt \lineskip-.5pt
   \ialign{$\m@th#1\hfil##\hfil$\crcr#2\crcr\sim\crcr}}}

\begin{document}

\title{
Competing Orders and Disorder-induced Insulator to Metal Transition 
in Manganites
}
\author{
Yukitoshi Motome$^{1*}$, 
Nobuo Furukawa$^2$, and Naoto Nagaosa$^{1,3,4}$
}
\affiliation{
$^{1}$Tokura Spin SuperStructure Project, ERATO,
Japan Science and Technology Corporation,
c/o National Institute of Advanced Industrial Science and Technology,
Tsukuba Central 4, 1-1-1 Higashi, Tsukuba, Ibaraki 305-8562, Japan
\\
$^{2}$Department of Physics, Aoyama Gakuin University, 
5-10-1 Fuchinobe, Sagamihara, Kanagawa 229-8558, Japan
\\
$^{3}$CREST, Department of Applied Physics, University of Tokyo,
7-3-1 Hongo, Bunkyo-ku, Tokyo 113-8656, Japan
\\
$^{4}$Correlated Electron Research Center, AIST,
Tsukuba Central 4, 1-1-1 Higashi, Tsukuba, Ibaraki 305-8562, Japan
}
\date{\today}

\begin{abstract}
Effects of disorder on the two competing phases, i.e.,  
the ferromagnetic metal and the commensurate charge/lattice 
ordered insulator, are studied by Monte Carlo simulation. 
The disorder suppresses the charge/lattice 
ordering more strongly than the ferromagnetic order, 
driving the commensurate insulator to the ferromagnetic metal 
near the phase boundary in the pure case. 
Above the ferromagnetic transition temperature, 
on the contrary, the disorder makes the system more insulating, 
which might cause an enhanced colossal magnetoresistance 
as observed in the half-doped or Cr-substituted manganites. 
No indication of the percolation or the cluster formation is found, 
and there remain the charge/lattice fluctuations instead 
which are enhanced toward the transition temperature. 
\end{abstract}


\maketitle
%%%%%%%%%%%%%%%%%%% Introduction
The mechanism of the colossal magnetoresistance (CMR) in manganites 
is one of the central issues in the physics of 
the strongly correlated electronic systems 
\cite{Tokura2000,Dagotto2001}. 
By applying the external magnetic field of a few Tesla, 
the resistivity changes of the order of $10^4$-$10^6$. 
Although the double-exchange (DE) mechanism gives a reasonable description 
for the negative magnetoresistance 
\cite{Furukawa1999}, 
it is not enough to explain such a huge response. 
The first proposal for the mechanism of the CMR was based on 
the transition from small to large polaron driven by the magnetic field 
\cite{Millis1995}. 
The idea is that the effective bandwidth controlled 
by the DE interaction is enhanced by the spin ferromagnetic 
alignment under the external magnetic field, 
which leads to the crossover from small to large Jahn-Teller polaron. 
However, the change of the effective bandwidth 
is rather small, and this mechanism does not lead to such a huge effect. 
Thus, this single-particle picture is not satisfactory 
to explain the CMR. 

Instead, recently, it has been recognized that the CMR is 
a collective phenomenon due to the many-body correlation. 
One of the present authors proposed 
a new mechanism of CMR based on the multicritical fluctuation 
between the ferromagnetic metal (FM) and the charge-ordered insulator 
(COI) occurring near $x=0.5$ ($x$: hole concentration) 
\cite{MurakamiPREPRINT}. 
There, the enhanced fluctuations near the multicritical point 
trigger a giant response to the external magnetic field. 
An evidence for this multicritical scenario is 
the scaling law for the magnetization curve (Arrott plot). 
However, this scaling analysis holds only for some class of materials, 
while the others do not show a good agreement. 

In the latter class of materials, 
the disorder appears to play an important role. 
Especially the spatially inhomogeneous structure has been observed 
experimentally \cite{Mori1998} and the percolation mechanism of 
the CMR is phenomenologically proposed based on this observation 
\cite{Moreo1999}. 
Namely it is assumed that the resistivity is determined 
by the percolating conduction paths, 
which are so sensitive to the pattern of 
the coexisting metallic and insulating regions 
under the influence of the disorder. 
The external magnetic field changes this pattern, 
and hence causes a large change of the resistivity. 

\begin{figure}
\includegraphics[width=6.5cm]{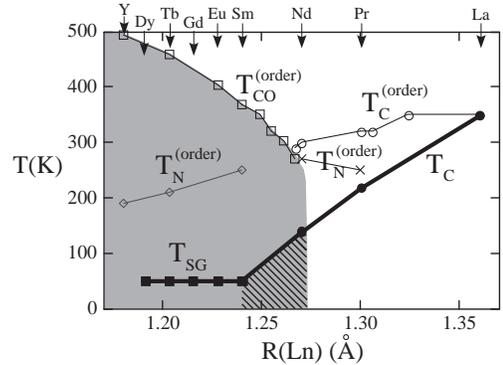}
\caption{
Phase diagram of {\it Ln}$_{1/2}$Ba$_{1/2}$MnO$_3$ 
from Ref.~[8]. 
$T_{\rm C}$, $T_{\rm CO}$, $T_{\rm N}$, and $T_{\rm SG}$ denote 
the ferromagnetic, charge-ordering, antiferromagnetic, and 
spin-glass transition temperatures, respectively. 
The subscript `(order)' are for the ({\it Ln},Ba)-ordered materials. 
The shaded area shows the charge-ordered insulating state 
in the ordered materials. 
The hatched area represents the disorder-induced ferromagnetic phase. 
} 
\label{fig:phase_exp} 
\end{figure}

Recent experiments in the half-doped manganites 
{\it Ln}$_{1/2}$Ba$_{1/2}$MnO$_3$, where {\it Ln} is a rare-earth element, 
have elucidated more explicitly the importance of the disorder
\cite{AkahoshiPREPRINT,Nakajima2002}. 
In these compounds, the strength of disorder due to the solid solution of 
({\it Ln},Ba) ions can be tuned by careful treatment of the synthesis 
\cite{Millange1998}. 
Figure~\ref{fig:phase_exp} shows the experimental phase diagram 
reported in Ref.~\cite{AkahoshiPREPRINT}. 
In the `clean' limit where {\it Ln} and Ba ions form 
a periodic layered structure,
the phase diagram shows a multicritical behavior
where FM and COI compete with each other. 
On the other hand, in the `disordered' case where 
the arrangement of ({\it Ln},Ba) ions is random, 
the phase diagram changes in a very asymmetrical manner: 
FM is suppressed but survives at finite temperatures, while COI disappears 
and instead some glassy state is realized at low temperatures. 
This leads to a nontrivial regime where the disorder induces 
the transition from COI to FM (hatched area in Fig.~\ref{fig:phase_exp}).
The enhanced CMR effect is observed above $T_{\rm C}$ in this regime 
\cite{AkahoshiPREPRINT}.
Similar phenomena are observed also in 
{\it Ln}$_{0.55}$(Ca,Sr)$_{0.45}$MnO$_3$ \cite{TomiokaPREPRINT} and 
in {\it Ln}$_{0.5}$Ca$_{0.5}$MnO$_3$ with the Cr substitution 
of Mn \cite{Barnabe1997}. 

Therefore, to understand the mechanism of the CMR, 
it is crucial to clarify the effects of the quenched disorder on the 
competition between the FM and CO, and on 
the metal-insulator transition, which we undertake in this letter.
Employing the unbiased numerical method to treat large fluctuations properly, 
we reveal that these two orders show
contrastive responses to the random potential 
which indeed leads to the disorder-induced transition from COI to FM. 
Through the systematic study of density of states,
conductivity, and fluctuations, 
we elucidate the mechanism of the enhanced CMR effect in this regime.

%%%%%%%%%%%%%%%%%%% Model
We consider a minimal model to describe a competition 
between FM of the DE origin and 
COI concomitant with the lattice distortion, 
whose Hamiltonian reads 
\begin{eqnarray}
&H& = -t \sum_{\langle ij \rangle \sigma} (c_{i\sigma}^\dagger c_{j\sigma} 
+ {\rm h.c.}) - J_{\rm H} \sum_i \sigma_i^z S_i 
\nonumber \\
&-&g \sum_i n_i Q_i 
+ \frac12 \sum_i Q_i^2 + \frac{\lambda}{2} \sum_{\langle ij \rangle} Q_i Q_j
+ \sum_i \varepsilon_i n_i.
\label{eq:H}
\end{eqnarray}
Here, the first two terms represent a simplified DE model 
which consists of the electrons' nearest-neighbor hopping $t$ and 
the Hund's-rule coupling $J_{\rm H}$. 
We consider the Ising-type Hund's-rule coupling 
in the limit of $J_{\rm H} \rightarrow \infty$ for simplicity, 
which retains an essential physics of the DE mechanism 
\cite{Motome2001}. 
The third term describes the electron-phonon coupling. 
Here, we consider the classical phonons 
which couple to the electron density (breathing mode). 
The next two terms are the elastic energy of phonons. 
We take account of the cooperative aspect of the lattice distortion 
by the fifth term which reflects the fact that 
a displacement of an oxygen affects both MnO$_6$ tetrahedra 
connected by the oxygen. 
The last term incorporates the random on-site potential energy $\varepsilon_i$ 
which couples to the electron density 
$n_i = \sum_\sigma c_{i\sigma}^\dagger c_{i\sigma}$. 
The alloying effect of {\it A}-site ions in {\it A}MnO$_3$ 
and the Cr substitution into the Mn site
cause the structural and electrostatic disorder, 
which disturbs the on-site potential 
through the Madelung energy. 
Since large fluctuations in the competition between different orders are crucial 
in this study, we employ the Monte Carlo (MC) method in which the 
configurations of $\{ S_i \}$ and $\{ Q_i \}$ are stochastically sampled 
\cite{Yunoki1998a}. 
In the presence of the disorder, we take the random average 
for different configurations of $\{ \varepsilon_i \}$. 

We consider model (\ref{eq:H}) on the 2D square lattice 
in the half-doped case with $x=0.5$ 
($0.5$ electron per site on average). 
We set the half-bandwidth $W=4t$ as an energy unit, and 
take $\lambda=0.1$ throughout this paper. 
A similar model was studied 
in a different context for the low-$x$ regime 
in the absence of the disorder 
\cite{Verges2002}. 
We consider the binary-type distribution of the random potential, 
$\varepsilon_i = \pm \Delta$.  
We apply the systematic analysis on the finite-size effect 
by using the series of $L \times L$-site systems with $L=4,6,8$
to distinguish the long-range ordering from short-range correlations. 
The numerical details will be reported elsewhere 
\cite{MotomePREPARATION}.

\begin{figure}
\includegraphics[width=7.5cm]{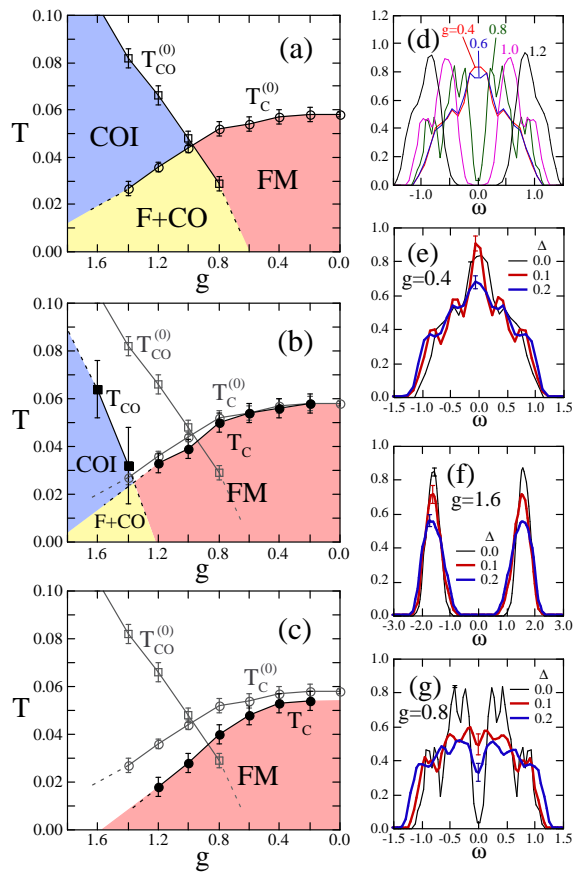}
\caption{
(a)-(c) Phase diagram of model (\ref{eq:H}) 
for different strength of the random potential
(a) $\Delta=0.0$, (b) $\Delta=0.1$, and (c) $\Delta=0.2$. 
The lines are guides for the eyes. 
The gray data and lines in (b) and (c) are the results in (a) for comparison. 
(d)-(g) Density of states at $T=0.016$ for $L=8$ clusters: 
(d) $\Delta=0.0$ for various $g$; 
(e) $g=0.4$, (f) $g=1.6$, and (g) $g=0.8$ by varying $\Delta$. 
The Fermi energy is at $\omega=0$. 
Only the typical error is shown for each case. 
} 
\label{fig:phase} 
\end{figure}

%%%%%%%%%%%%%%%%%%% Results and Discussions
Figure~\ref{fig:phase} (a) shows the phase diagram of model (\ref{eq:H}) 
in the absence of the disorder ($\Delta=0$). 
Note that the abscissa is reversed for comparison with experimental results. 
At $g=0$, the electronic part of model (\ref{eq:H}) 
becomes a pure DE model which exhibits 
the FM state below $T_{\rm C}^{(0)}$ 
(the superscript $(0)$ represents the case of $\Delta=0$). 
As $g$ increases, $T_{\rm C}^{(0)}$ decreases and 
the checkerboard-type COI concomitant with the cooperative lattice distortion 
appears below $T_{\rm CO}^{(0)}$. 
In the intermediate region where $T_{\rm C}^{(0)}$ and $T_{\rm CO}^{(0)}$ 
intersect, we have the phase with both the ferromagnetism and 
the charge/lattice ordering (F+CO phase). 
Thus, the phase diagram exhibits the tetracritical behavior
where four different phases meet at one point. 
The density of states (DOS) shown in Fig.~\ref{fig:phase} (d) indicates that 
the phases below $T_{\rm CO}^{(0)}$ including the F+CO phase 
have a finite energy gap. 
This tetracritical behavior can be turned into 
the bicritical one when we include a competing term 
between the ferromagnetism and the charge/lattice ordering by hand 
\cite{MotomePREPARATION}, 
and the following results on the disorder effects are 
qualitatively similar in these multicritical phenomena. 

The emergence of the F+CO phase is due to the robustness of 
the DE ferromagnetic interaction even in the insulating state. 
The DE interaction is effective
if the electron wavefunction spreads, at least, for a finite spatial range. 
We note that this phase is also predicted in more realistic DE model for 
manganites \cite{AliagaPREPRINT,Yunoki1998b}, and 
is indeed observed in recent experiments \cite{Loudon2002}. 

When we switch on the random potential, the phase diagram shows a drastic 
change. 
By introducing weak disorder with $\Delta=0.1$ (Fig.~\ref{fig:phase} (b)), 
$T_{\rm CO}$ is suppressed remarkably, 
whereas $T_{\rm C}$ is lowered only slightly. 
Accordingly, the tetracritical point shifts 
toward larger $g$ and lower $T$. 
For stronger disorder with $\Delta=0.2$ (Fig.~\ref{fig:phase} (c)), 
$T_{\rm CO}$ disappears down to the lowest temperatures 
in the present MC calculations ($T \simge 0.008$). 
This clearly shows the fragility of COI
against the random potential compared to the DE ferromagnetism 
\cite{AliagaPREPRINT}. 
As a result of this asymmetric effect, 
there appears the regime where COI 
is turned into FM by introducing the random potential, 
which is not clearly found in the previous study 
\cite{AliagaPREPRINT}. 
These features are in good agreement with the experimental results 
\cite{AkahoshiPREPRINT,TomiokaPREPRINT}. 

The DOS gives insights on the nature of these disorder effects. 
As shown in Fig.~\ref{fig:phase} (e) and (f), 
DOS in both the small-$g$ FM regime and the large-$g$ COI regime are
not much affected by the random potential. 
Especially, Fig.~\ref{fig:phase} (f) indicates that 
the energy gap in the large-$g$ regime persists 
even when the charge/lattice long-range order is destroyed 
by the disorder. 
This means that a short-range correlation remains and 
the local lattice distortion persists to open the gap. 
Therefore, the collapse of the charge/lattice ordering 
is not due to the disappearance of 
the amplitude of the lattice distortion 
but due to the disturbance of the phase of the commensurate ordering
with the wavenumber ($\pi,\pi$). 
The random potential acts as the `random field' to the CO order parameter, 
and its phase coherence is lost.
On the contrary, the ferromagnetic order is robust against such pinning 
since $T_{\rm C}$ is essentially determined by the kinetic energy of electrons. 
These give a comprehensive understanding 
of the asymmetrical effect of the disorder on the phase diagram. 

The most remarkable behavior of DOS is obtained near the phase boundary 
between the FM state and the F+CO insulating state. 
In the F+CO phase but close to the phase boundary at $\Delta=0$, 
as shown in Fig.~\ref{fig:phase} (g), 
a small but clear gap in DOS below $T_{\rm CO}^{(0)}$ is rapidly collapsed 
by introducing the disorder and the system appears to be metallic. 
This indicates that the random potential easily destroys COI and induces FM 
in the proximity of the metal-insulator transition. 
This result also agrees well with experimental results 
\cite{AkahoshiPREPRINT,TomiokaPREPRINT}, 
which strongly suggests that the simple model (\ref{eq:H}) captures 
the essential physics of this competition between FM and CO in manganites. 

\begin{figure}
\includegraphics[width=7.5cm]{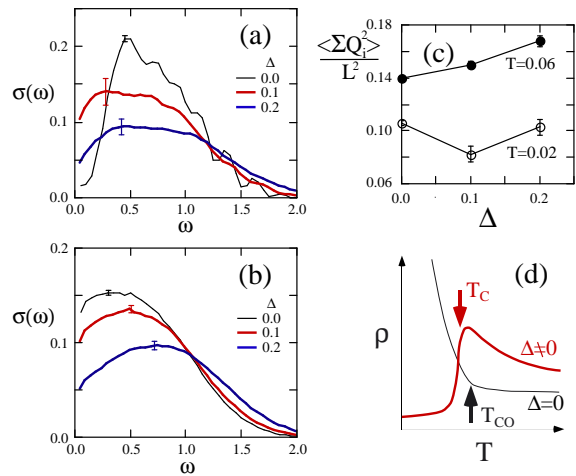}
\caption{
Optical conductivity for $g=0.8$ at (a) $T=0.02$ and (b) $T=0.06$ ($L=6$). 
(c) Average of the square of the lattice distortion ($L=8$). 
See the text for details. 
(d) Schematic picture of the temperature dependence of 
the resistivity.
} 
\label{fig:conductivity} 
\end{figure}

Above $T_{\rm C}$ of the disorder-induced metallic state, 
the CMR is found to be much enhanced experimentally
\cite{AkahoshiPREPRINT,TomiokaPREPRINT}. 
There, the system is insulating and 
shows a large drop of the resistivity near $T=T_{\rm C}$
toward the low-$T$ metallic phase. 
Our results are satisfactorily consistent with this behavior: 
Figures~\ref{fig:conductivity} (a) and (b) show the optical conductivity
$\sigma(\omega)$ for $T < T_{\rm C}$ and $T > T_{\rm C}$, respectively, 
which are obtained by the Kubo formula. 
In the low-$T$ regime, $\sigma(\omega)$ shows a disorder-induced 
insulator to metal transition which is consistent with 
DOS in Fig.~\ref{fig:phase} (g). 
On the contrary, at the high-$T$ regime, the disorder substantially reduces
the low-energy weight and develops a quasi-gap structure in $\sigma(\omega)$, 
namely, it makes the system more insulating.
These tendencies are also indicated in the magnitude of the lattice distortion, 
$\langle \sum_i Q_i^2 \rangle / L^2$, as shown in Fig.~\ref{fig:conductivity} (c),
which correlates with the localization of electrons. 
The distortion increases for $T > T_{\rm C}$ 
while it decreases for $T < T_{\rm C}$ 
when the random potential is introduced. 
These contrastive effects of the disorder suggest 
the temperature dependence of the resistivity $\rho$ 
as schematically drawn in Fig.~\ref{fig:conductivity} (d). 
In the absence of the disorder, the system causes the COI transition 
at $T_{\rm CO}^{(0)}$, where $\rho$ shows a sharp upturn. 
The disorder enhances the insulating nature in the high-$T$ regime, 
hence $\rho$ becomes larger 
while it does not show any anomaly corresponding to the COI transition. 
Below $T_{\rm C}$, the disorder induces FM, therefore
$\rho$ should show a large drop at $T \sim T_{\rm C}$ 
as shown in the figure. 
The large change of $\rho$ potentially leads to a huge response 
to the external magnetic field as is indeed observed in the CMR manganites 
\cite{AkahoshiPREPRINT,TomiokaPREPRINT}. 
Thus, the origin of the enhanced CMR can be attributed to 
the contrastive influence of the disorder
in low-$T$ and high-$T$ regimes near the metal-insulator phase boundary. 

\begin{figure}
\includegraphics[width=8.5cm]{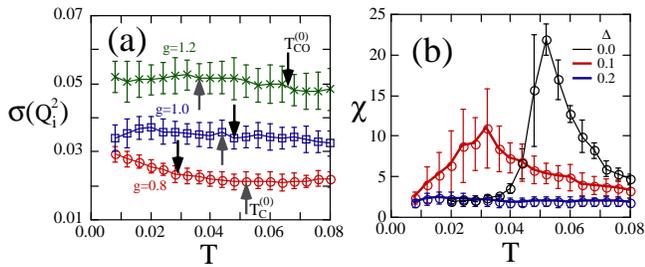}
\caption{
(a) Temperature dependence of the standard deviation of 
the square of the local lattice distortion at $\Delta=0.2$. 
Transition temperatures for FM (COI) in the absence of the disorder 
are indicated by the gray (black) arrows. 
(b) Susceptibility for the ($\pi,\pi$) lattice ordering at $g=1.0$. 
Both are for $L=8$ clusters.
} 
\label{fig:rough,kai} 
\end{figure}

We discuss here this insulating state for $T > T_{\rm C}$ in more detail. 
In the percolation scenario for the CMR \cite{Moreo1999}, 
the system is supposed to be a coexisting mixture of 
the metallic and the insulating clusters below 
the FM or COI transition temperatures in the absence of the disorder, 
$T_{\rm C}^{(0)}$ and $T_{\rm CO}^{(0)}$, respectively. 
However, our simulation does not show any clear indication of this coexistence. 
Figure~\ref{fig:rough,kai} (a) shows the standard deviation of 
the square of the lattice distortion 
$\langle Q_i^2 \rangle$ among the sites 
as a function of the temperature. 
The standard deviation changes monotonously 
and remains small (compare to the values in Fig.~\ref{fig:conductivity} (c)). 
This is contrary to the percolation picture 
in which we expect an enhancement for $T < T_{\rm C}^{(0)}$ or 
$T_{\rm CO}^{(0)}$  
due to the formation of the metallic (smaller $\langle Q_i^2 \rangle$) and 
the insulating (larger $\langle Q_i^2 \rangle$) clusters. 
Instead of the cluster formation, we find fluctuations or 
short-range correlations of the charge/lattice ordering. 
Figure~\ref{fig:rough,kai} (b) shows the susceptibility for the ($\pi,\pi$) 
lattice ordering. 
In the absence of the disorder, the susceptibility shows 
a sharp cusp at $T=T_{\rm CO}^{(0)}$, which leads to the divergence 
in the thermodynamic limit $L \rightarrow \infty$. 
Although the disorder destroys COI and makes the cusp obscure, 
the fluctuation remains finite and is enhanced as $T \rightarrow T_{\rm C}$. 
The contrastive behavior between above and below $T_{\rm C}$ 
in the disordered case is in accord with 
this temperature dependence of $\chi$. Namely 
the CO fluctuation is enhanced towards $T_{\rm C}$ from above while 
it is suppressed at the lowest temperature. This could be regarded as
the reminiscence of the multicritical phenomenon in the pure case.
This is also in good agreement with the diffuse scattering experiments 
which indicate a large fluctuation of the 
charge/lattice ordering above $T_{\rm C}$ 
\cite{TomiokaPREPRINT,Shimomura1999}. 
We note that this fluctuation is suppressed as $\Delta$ increases, 
which may account for the deviation from the scaling law 
\cite{MurakamiPREPRINT}. 
Therefore, our results indicate that the phase above $T_{\rm C}$ is not 
likely the inhomogeneous mixture of the static clusters
but a rather homogenous state with the thermodynamic
charge/lattice fluctuations.

%%%%%%%%%%%%%%%%%%% Acknowledgment

The authors acknowledge Y. Tokura, Y. Tomioka, and E. Dagotto 
for fruitful discussions. 
This work is supported by Grant-in-Aids from the Ministry of Education,  
Culture, Sports, Science, and Technology.

\noindent
$^*$ Present address: RIKEN (The Institute of Physical and Chemical Research), 
2-1 Hirosawa, Wako, Saitama 351-0198, Japan.

%%%%%%%%%%%%%%%%%%% References

%%%%% FIGURES

\end{document}